\documentclass{epl}
\usepackage[centertags]{amsmath}
\usepackage{epsfig}
\usepackage{psfrag}

\title{Numerical entropy and phason elastic constants of plane random
tilings with any $2D$-fold symmetry}
\shorttitle{Entropy and phason elastic constants of tilings with $2D$-fold 
symmetry}
\author{N. Destainville\inst{1}}

\institute{                    
  \inst{1} Laboratoire de Physique Th\'eorique, IRSAMC - UMR CNRS/UPS 5152 \\
Universit\'e Paul Sabatier,
31062 Toulouse Cedex, France.
}

\def\openone{{\leavevmode\hbox{\small1\kern-3.55pt\normalsize1}}}

\def\sbar{\overline{\sigma}}

\def\EC{{\mathcal E}}
\def\vect#1{\mbox{\boldmath{$#1$}}}

\def\Z{\mbox{\boldmath{$Z$}}}
\def\R{\mbox{\boldmath{$R$}}}

\begin{document}

\maketitle

\begin{abstract}
We perform Transition matrix Monte Carlo simulations to evaluate the
entropy of rhombus tilings with fixed polygonal boundaries and
$2D$-fold rotational symmetry. We estimate the large-size limit of
this entropy for $D=4$ to 10. We confirm analytic predictions of
N. Destainville {\em et al.}, {\em J. Stat. Phys.} {\bf 120}, 799
(2005) and M. Widom {\em et al.}, {\em J. Stat. Phys.} {\bf 120}, 837
(2005), in particular that the large size and large $D$ limits
commute, and that entropy becomes insensible to size, phason strain
and boundary conditions at large $D$. We are able to infer finite
$D$ and finite size scalings of entropy.  We also show that
phason elastic constants can be estimated for any $D$ by measuring the
relevant perpendicular space fluctuations.
\end{abstract}

Random tiling models~\cite{henley91,elser85} have been intensively
studied since the discovery of quasicrystals in 1984, because they are
good paradigmatic models of quasicrystals. These metallic compounds
exhibit exotic symmetries ({\em e.g.}  icosahedral) which are
classically forbidden by crystallographic rules. This is accounted for
by the existence of underlying quasiperiodic or random tilings. When
tiles are decorated in some manner by atoms, these tilings become
excellent candidates for modeling real quasycristalline
compounds~\cite{AlNiCo}. As compared to perfectly quasiperiodic
tilings, some specific degrees of freedom, the so-called {\em phason
flips}, are activated in random tilings, giving access to a large
number of microscopic configurations. The number of configurations
grows exponentially with the system size in contrast to perfectly
quasiperiodic tilings where it only grows polynomially.  The resulting
configurational entropy lowers the free energy as compared to
competing crystalline phases.  Despite their random character, random
tilings still display the required macroscopic point symmetries in
their Fourier spectra. They are as good candidates as perfectly
quasiperiodic tilings for modeling
quasicrystals~\cite{henley91,elser85}. The statistical mechanics of
random tilings is of central interest for quasi-crystallography.  But
the calculation of the thermodynamical observables of interest in
random tiling models has turned out to be a formidable task. Even the
calculation of configurational entropy in the case of {\em maximally}
random tilings where all tilings have the same energy is a notoriously
difficult problem. Very few models are exactly
solvable~\cite{soluble1,soluble2,soluble2b,soluble3,soluble3b,soluble4},
and a large majority of calculations rely on numerical simulations. In
Refs.~\cite{largecodimI,largecodimII}, an original analytic mean-field
theory for plane rhombus tilings with $2D$-fold symmetry, in the large
$D$ limit, was proposed. Its ultimate goal is to derive valuable
results on finite $D$ tilings by estimating finite $D$ corrections to
the infinite $D$ limit. The present study intends to support
numerically the analytic findings of this promising approach.  We
propose a calculation of the entropy per tile of rhombus tilings based
on a Transition matrix Monte Carlo (TMMC)
technique~\cite{arctic3D}. We get estimates of the large size limit of
the entropy for $D \leq 10$ (Table~\ref{large:p:limits}).
Furthermore, we calculate the same entropy for large $D$ tilings
filling $2D$-gons of finite side length $p$. We confirm the
theoretical predictions of Refs.~\cite{largecodimI,largecodimII}:
large $D$ and large $p$ limits commute. We also explore the large $D$
behavior of a phason elastic constant by measuring the relevant
perpendicular space fluctuations. We show that it decays like $1/D$.

We study tilings that fill centro-symmetric $2D$-gons
(Fig.~\ref{exs}). Such fixed boundaries have a strong influence on
tilings at finite $D$ (see~\cite{largecodimII,arctic3D} and references
therein). Their most spectacular consequence is that the entropy per
tile of fixed-boundary tilings is strictly smaller than its free- or
periodic-boundary counterpart. It was also one of the goals of
Ref.~\cite{largecodimII}, where these questions are thoroughly
discussed, to explore what this difference between fixed and
free boundary conditions becomes in the large $D$ limit. It was demonstrated
there that, in this limit, all boundary conditions become equivalent.
More generally, it was proved that under very weak conditions, the
entropy per tile of large $D$ tilings is independent of their size,
shape, tile fractions and boundary conditions. The ``universal''
entropy per tile was estimated by a mean-field approach and
numerically calculated~: $\sigma_{\infty} \simeq 0.5676 \pm 0.0001$.
It was also argued that some phason elastic constants associated with
given strain modes vanish at large $D$. Our goal here is to explore
numerically these predictions in a larger variety of situations.

The tiles considered in this paper are lozenges of unitary side length
and of angles multiple of $\pi/D$.  Their edges are collinear to the
$D$ vectors $\vect{e}^{\parallel}_{\alpha}$, $\alpha=0,\ldots,D-1$
which make angles $\alpha \pi/D$ with an arbitrary direction. In the
celebrated {\em cut-and-project} framework (see~\cite{henley91}),
rhombic tiles are considered as the projections onto the
2-dimen\-sional space of the $2$-dimensional faces of a hypercube in a
space of dimension $D$ denoted by $H$. Conversely, any tiling in the
bidimensional space can be ``lifted'' to a bidimensional continuous
membrane embedded in the $\Z^D$ lattice. When this membrane is
projected back into the 2-dimensional space, the facets of the
$D$-dimensional lattice project precisely onto the rhombic tiles. The
normalized basis vectors of $H$ are denoted by
$\vect{e}_{\alpha}$. The $\vect{e}^{\parallel}_{\alpha}$ are the
normalized projections of the $\vect{e}_{\alpha}$~\cite{largecodimI}.
The physical 2-dimensional space is usually called the {\em parallel
space} and it is denoted here by $\EC^{\parallel}$. The
$(D-2)$-dimensional space in $H$ perpendicular to $\EC^{\parallel}$ is
called the {\em perpendicular space}.  It is denoted by
$\EC^{\perp}$. The normalized projections of the $\vect{e}_{\alpha}$
in $\EC^{\perp}$ are the $\vect{e}^{\perp}_{\alpha}$. The $2D$-gons
filled by such tiles in the present publication have edges parallel to
the $\vect{e}^{\parallel}_{\alpha}$.  Their integral side lengths are
denoted by $p_{\alpha}$. When all $p_{\alpha}$ are equal to the same
$p$, the tilings are said here to be {\em diagonal, of side
$p$}. Local moves called single flips are usually used to explore
configuration spaces of rhombus tilings.  In dimension 2, they consist of the
exchange of three tiles that locally fill a small elementary hexagon
(see Fig.~\ref{exs}, left). Such flips connect configuration spaces in
dimension 2.

\section {Transition matrix Monte Carlo technique} 

\begin{figure}[ht]
\begin{center}
\resizebox{!}{3cm}{
\includegraphics{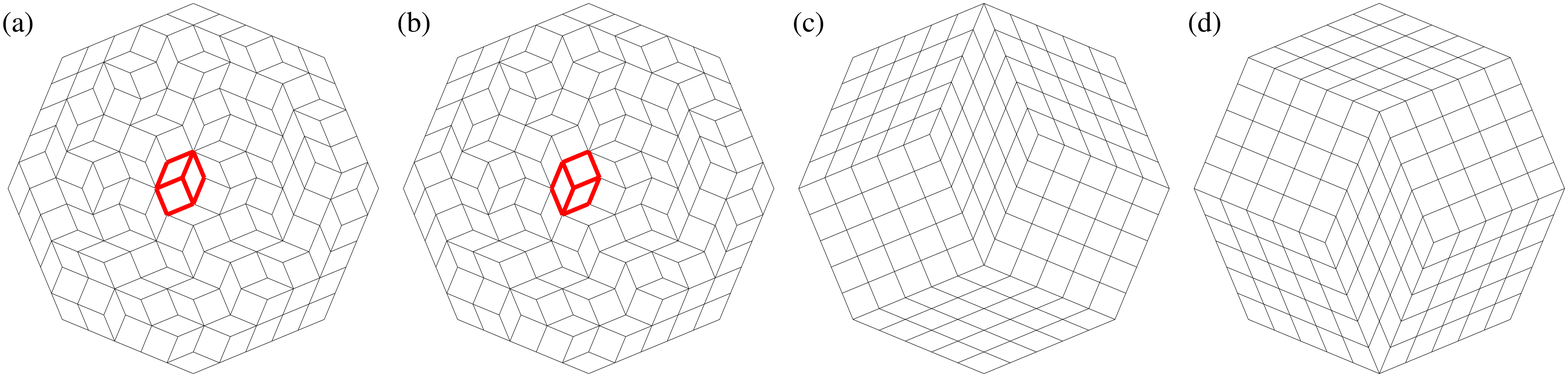}
} 
\caption{Several examples of tilings of side
$p=5$ with $D=4$. (a,b): two typical random tilings differing by a single
flip in the emphasized hexagon. (c,d): Tilings with minimal energy
$E_{min}$ (c) and maximal energy $E_{max}$ (d). Tilings (a) and (b)
have energies close to $E_{mid}=(E_{min}+E_{max})/2$. \label{exs}}
\end{center}
\end{figure}

The previous estimates of the large $D$ entropy~\cite{largecodimII}
relied on an efficient numerical approach in the $p=1$ case.  It
cannot, however, be applied to values of $p > 1$ of interest in the
present paper, without inextricable technical complications. We chose
to apply here a Transition matrix Monte Carlo
technique~\cite{arctic3D}. It is a variant of the transition matrix
method~\cite{TMMC1}. It uses a standard Metropolis Monte Carlo
sampling to construct a numerical approximation to the transition
matrix. The density of states is calculated from this transition
matrix and provides the total number of configurations. The energy $E$
used here in order to implement the Monte Carlo algorithm has no
particular physical meaning as far as quasicrystals are concerned. It
is related to the structure of the configuration space in terms of
partial ordered sets (posets) theory~\cite{Desoutter05}. It involves
the description of rhombus tilings with polygonal boundaries by means
of {\em generalized partitions}, itself related to the cut-and-project
technique. Its complete construction is detailed in
Refs.~\cite{arctic3D} and~\cite{Desoutter05} (where it is equal to the
so-called ``rank function''). There is a unique tiling $t_{min}$
(resp. $t_{max}$) with minimal energy $E_{min}$ (resp. maximal energy
$E_{max}$) (see Fig.~\ref{exs}, bottom). The energy measures the
distance from a given tiling $t$ to $t_{min}$.  A single flip as
described above increases or decreases the energy by a single unit,
according to whether it increases or decreases the distance from
$t_{min}$. The structure of graded poset~\cite{Desoutter05} ensures
that the energy variation between any two tilings does not depend on
the choice of the path of flips between these tilings. We find bellow
that the energy of most tilings is close to
$E_{mid}=(E_{min}+E_{max})/2$. The density of states is symmetric with
respect to $E_{mid}$, because of the geometric centro-symmetry of the
problem.

In the Metropolis Monte Carlo algorithm, at temperature $T\geq0$, a
randomly chosen vertex is tried to be flipped if it is flippable. This
change is accepted if it lowers the energy. If by contrast the change
raises the energy, it is accepted with probability $\exp(-\Delta
E/T)$, where $\Delta E$ is the energy variation {\em if the change was
accepted}. At low $T$, tilings are close to $t_{min}$. At large $T$,
they are in the maximum entropy region, close to $E_{mid}$. It will
also be useful to take negative values of $T$ after adapting suitably
the previous transition probabilities. Now we use the fact
that the energy variation associated with a single flip cannot be
larger than $\pm 1$. As in Ref.~\cite{arctic3D}, we denote by
$n_{\pm}(t)$ the number of tilings that can be reached from a given
tiling $t$ by single upwards or downwards flips. For any given tiling
$t$, the numerical calculation of $n_{\pm}(t)$ is easy and exact. If
$N_V$ is the number of tiling vertices, then $n_+(t)+n_{-}(t) \leq
N_V$. Remind that $N_V$ is tiling-independent. In particular, it is
unchanged through a single flip.

Now we can define the {\em transition matrix} $\omega(E,E')$ of
dimension  $E_{max}-E_{min}+1$. It is equal to 0 everywhere except
on its diagonal and two off-diagonals:
\begin{equation}
\omega(E,E\pm 1) = \frac{1}{W(E)} \sum_{t \in P(E)}
\frac{n_{\pm}(t)}{N_V},
\label{omega1}
\end{equation}
where $P(E)$ is the set of tilings of energy $E$ and $W(E)$ is their
number. In addition $\omega(E,E) = 1 - \omega(E,E-1) - \omega(E,E+1)$.
The matrix $\omega$ is not known and it is the goal of our Monte Carlo
algorithm to estimate it by sampling the energy levels $E$ and
estimating the exact mean~(\ref{omega1}) by an approximate average on
sampled configurations. The density of states can then be extracted as
follows: $\omega_+(E) W(E) = \omega_-(E+1) W(E+1)$ because the total
number of forward flips from energy $E$ to energy $E+1$ is exactly
equal to the total number of backward flips from energy $E+1$ to
energy $E$. This equation reads $W(E+1) = \omega_+(E)
W(E)/{\omega_-(E+1)}$.  It allows us to iteratively extract the $W(E)$
using uniqueness of the ground state $t_{min}$,
$W(E_{min})=1$. Finally, the total number of tilings is $Z=\sum_E
W(E)$, the configurational entropy is $S= \ln Z$, and the entropy per
tile is $\sbar_D(p)=S/N_T$, $N_T$ the number of tiles.
To ensure that all energy levels are (almost) uniformly visited, we
perform sweeps over temperature, for both positive and negative
temperatures. 
The minimum and maximum temperatures $T_{min}$ and $T_{max}$ must be
suitably adjusted to ensure that this distribution is as uniform as
possible in order to sample correctly all energy levels. We find that
the density of states $W(E)$ is nearly Gaussian, at least near its
maximum at $E_{mid}$~\cite{arctic3D}.

\section{Numerical results} 

We first focus on the {\em diagonal} case where $p_{\alpha} =
p$. Tilings are initialized at minimum energy $E_{min}$ which is
chosen to be equal to 0. The maximum energy can be calculated {\em a
priori}~\cite{arctic3D,Desoutter05}:
$E_{max} = {D \choose 3} \; p^3$.
We accumulate statistics $n_{\pm}(t)$ to estimate the transition
matrix. The degree to which symmetry of $S(E)$ with respect to $E_{mid}$ is
broken serves as an indicator of errors accumulated during the
iterative calculation. The residual $R=\ln W(E_{max})$,
that should ideally be equal to 0, reflects the cumulative errors in
$W(E)$. It can be demonstrated~\cite{arctic3D} that $R$ is a good
indicator of the uncertainty on $S$ at the end of the calculation. It
will provide error bars on $\sigma_D(p)$ below. In all cases, error
bars will be smaller than $10^{-4}$ (except for $D\geq50$ where
they will be smaller than $5.10^{-4}$).

\begin{figure}[ht]
\begin{center}
\ \psfig{figure=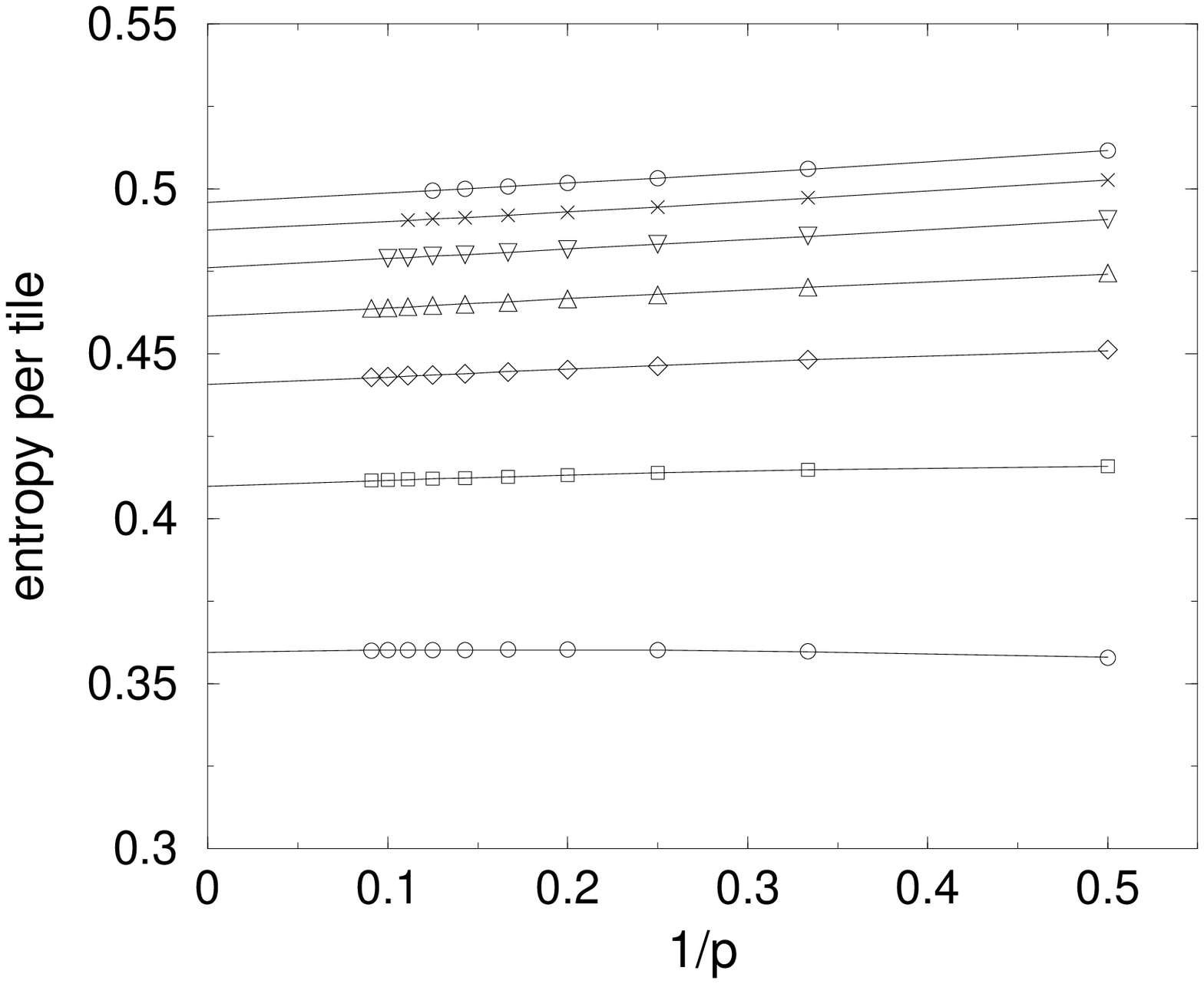,width=6cm}~\psfig{figure=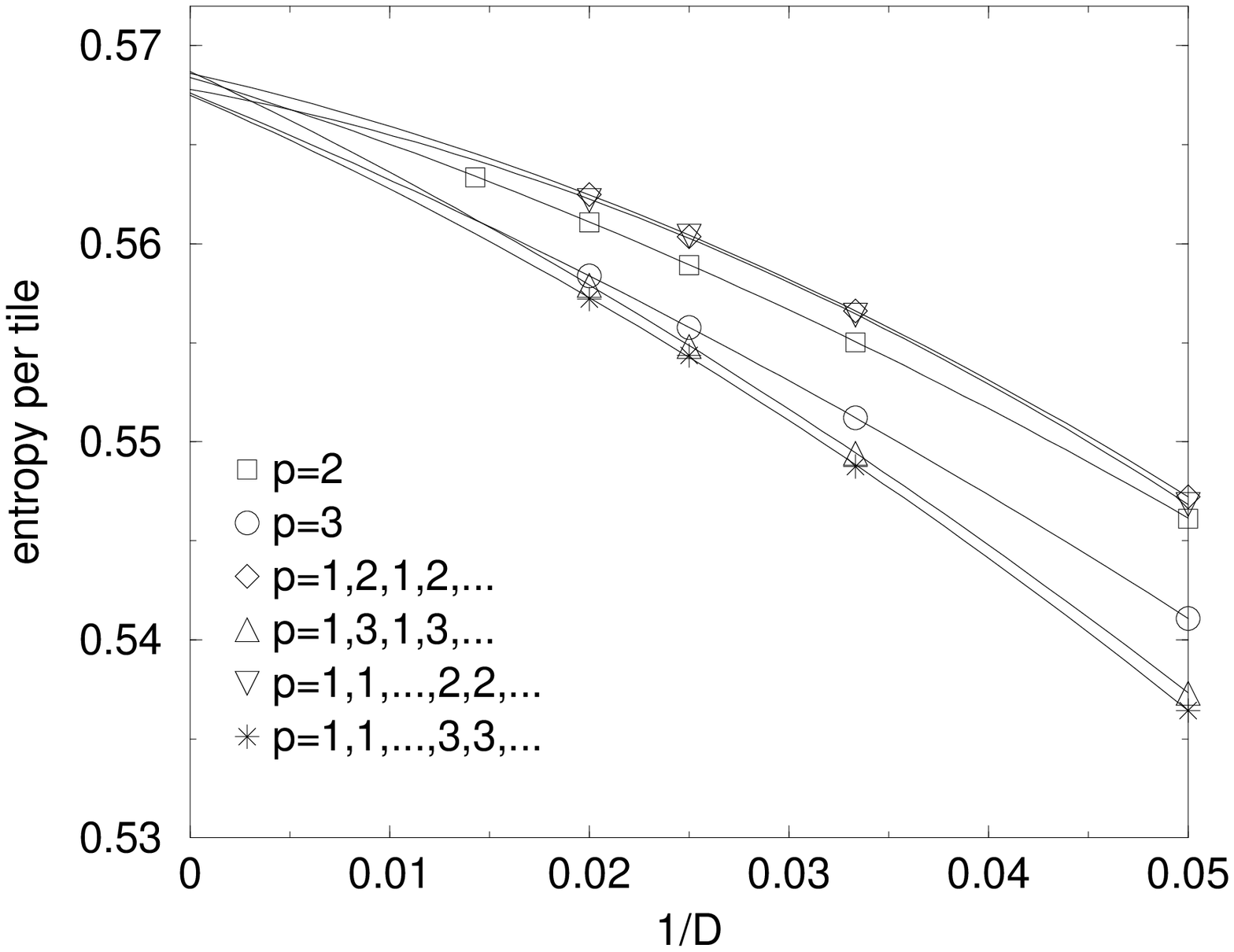, width=6.2cm} \
\end{center}
\caption{Left: Entropies per tile $\sbar_D(p)$ {\em vs} $1/p$ for
$D=4,5,6,7,8,9$ and $10$ from bottom to top (symbols) and 
second order fits (lines). Error bars are much smaller than symbol sizes
because all errors are smaller than $5 \; 10^{-4}$. 
Right: Entropies per tile {\em vs} $1/D$ in various situations
(symbols; see text) and their fits at order 2 in $1/D$ (lines). 
All plots converge towards the ``universal'' value $\sigma_{\infty} \simeq 
0.5676$ within our uncertainties.
\label{fitDfinite}}
\end{figure}

We begin with the infinite $p$ limit of finite $D$ entropies of
diagonal tilings with fixed polygonal boundaries: $\overline{\sigma}_D
= \lim_{p \rightarrow \infty} \sbar_D(p)$.  For $D=4-10$, we
extrapolate the large $p$ limit by fitting the numerical data obtained
{\em via} the previous algorithm. We choose a second order fit in
$1/p$. Contrary to the three-dimensional case~\cite{arctic3D}, we did
not observe that logarithmic corrections improved the quality of the
fit. Fig.~\ref{fitDfinite} (left) displays $\sbar_D(p)$ in
function of $1/p$ and the second order fits. The extracted large $p$
limits are listed in Tab.~\ref{large:p:limits}. We applied the same
procedure to the exactly solvable $D=3$ case (see~\cite{henley91}) and we
found that the method provides the exact expected value with a
relative precision smaller than 1\%. Therefore we anticipate that the
relative errors for larger values of $D$ are also smaller than 1\%.

\begin{table}[ht]
\caption{Large size limits of the entropy per tile.  
Relative errors are smaller than 1\% (see text).
\label{large:p:limits}}
$$
\begin{array}{|c|c|c|c|c|c|c|c|}
\hline
D & 4 & 5 & 6 & 7 & 8 & 9 & 10  \\
\hline
\; \sbar(D) \; & \; 0.360 \; & \; 0.410 \; & \; 
0.441 \; & \; 0.461 \; & \; 0.476 \; & \; 0.487 \; & \; 0.496 \; \\
\hline
\end{array}
$$
\end{table}

To test the predictions stated above, we also extrapolate the limit
$\sbar_{\infty} = \lim_{D \rightarrow \infty} \sbar_D = 0.563 \pm
0.006$ by fitting the finite $D$ values at order 2 in
$1/D$~\cite{largecodimII}. The error bar also corresponds to a
relative error of 1\%. We get $\sbar_{\infty}=\sigma_{\infty}$ within
error bars. This is a first confirmation of the predictions under
interest.  To verify them further, we also calculate the large $D$
limit for finite $p$. The case $p=1$ was solely considered in
Ref.~\cite{largecodimII}. We were able to perform more intensive
numerical calculations here. The entropy is extrapolated from finite
$D$ entropies for $D=20$ to 70 for $p=2$, and $D=20$ to 50 for
$p=3$. We find consistently $\sbar_{\infty} (p=2) = \sbar_{\infty}
(p=3) = 0.568 \pm 0.006$.  Then we consider boundary
strains~\cite{largecodimII}, with high frequency,
$(p_{\alpha})=(1,p,\ldots,1,p)$, with $p=2,3$, as well as low
frequency, $(p_{\alpha})=(1,\ldots,1,p,\ldots,p)$, with $D/2$ sides of
length 1 (resp. of length $p=2,3$). Extrapolating data for $20 \leq D
\leq 50$, we also find $\sbar_{\infty} (1,2,1,2,\ldots) =
\sbar_{\infty} (1,3,1,3,\ldots) = 0.569 \pm 0.006$ at high frequency
strain and $\sbar_{\infty} (1,\ldots,2,\ldots) = \sbar_{\infty}
(1,\ldots,3\ldots) = 0.568 \pm 0.006$ at low frequency. These
calculations, summarized in Fig.~\ref{fitDfinite} (right), also
support the predictions of the ``universality'' of the large $D$
limit~\cite{largecodimI,largecodimII}. They also confirm the
commutation of the large $p$ and large $D$ limits.

Now we examine finite $p$ and finite $D$ scalings. {\em A priori}, if
there are not logarithmic corrections, the Taylor expansion of $\sbar(D,p)$
at order 2 in powers of $1/p$ and $1/D$ reads
\begin{equation}
\label{generic}
\sbar(D,p) \simeq  \sbar_{\infty} + \frac{a}{p} + \frac{b}{D} +
\frac{c}{p^2} + \frac{d}{p\; D} + \frac{e}{D^2}.
\end{equation}
Since $\sbar(\infty,p) = \sbar_{\infty}$, $a=c=0$. Moreover, at fixed
$D=D_0$ and at first order in $p$, $\sbar(D_0,p) \simeq \left(
\sbar_{\infty} + \frac{b}{D_0} + \frac{e}{D_0^2}\right) +
\frac{d}{D_0} \; \frac{1}{p}$.  From the fit of the data displayed in
Tab.~\ref{large:p:limits}, we get $b \simeq -0.57$ and $e \approx
-1$. It is not possible to get a reliable estimate of $d$ by fitting
the available small $D_0$ data. Indeed, it appears that the slopes
$d/D_0$ do not have reached their asymptotic regime at $D=10$ because
they are still increasing with $D_0$. However, fitting these available
slopes at the second order in $1/D_0$ provides the rough estimate $d
\approx 0.2$.  In addition, at fixed $p=p_0$ and at first order in
$1/D$, Eq.~(\ref{generic}) becomes $\sbar(D,p_0) \simeq \sbar_{\infty}
+ ( b + d/p_0 )/D$.  Fitting the data for $p_0=2$ and $p_0=3$ gives
respectively $b+d/2 \simeq -0.31$ and $b+d/3 \simeq -0.41$, from which
we get $b \simeq -0.61$ and $d \simeq 0.60$. These values are
compatible with the previous ones.  In particular, both values of $b$
match very well, whereas they are extracted from the numerical data in
a different way: in the first case, we take the infinite $p$ limit
first, and in the second case, the infinite $D$ limit first. This
confirms {\em a posteriori} the general expression~(\ref{generic}). In
particular the two limits commute; There are probably no logarithmic
corrections in~(\ref{generic}).

\section{Phason elastic constants} 

Now we use the previous algorithm to estimate phason elastic
constants. Even though we are working on fixed-boundary tilings, our
goal is to measure elastic constants associated with {\em
free}-boundary ones. We use the notations and the conventions of
Ref.~\cite{largecodimI}. In particular, the normalizations must be
precisely prescribed: we define 3 scalar products in $H$,
$\EC^{\parallel}$ and $\EC^{\perp}$ denoted by $\langle . |
. \rangle_{V}$ where $V$ is the space under consideration. If
$\vect{x},\vect{y} \in \EC^{\perp}$, $\langle \vect{x} | \vect{y}
\rangle_{\EC^{\perp}} = s_{\perp}^2 \langle \vect{x} | \vect{y}
\rangle_{H}$ with $s_{\perp}^2 = D/(D-2)$~\cite{largecodimI}. In a
similar way, $s_{\parallel}^2 = D/2$. With these conventions, we get
$\| \vect{e}_{\alpha} \|_H = \| \vect{e}_{\alpha}^{\perp}
\|_{\EC^{\perp}} = \| \vect{e}_{\alpha}^{\parallel}
\|_{\EC^{\parallel}} = 1$, as desired.

We also employ here a useful variational principle (see
\cite{arctic3D} and references therein) to write the entropy in a
practical field-theoretic manner. After
coarse-graining~\cite{henley91}, large size tilings are represented by
regular height functions $\vect{\phi}: \R^2 \rightarrow \R^{D-2}$.
Such a ``macroscopic'' variable represents a large number
$N(\vect{\phi})$ of tilings, whose membrane representation in $\Z^D$
is equal, after coarse-graining, to the graph of $\vect{\phi}$. Their
micro-canonical entropy $\log(N(\vect{\phi}))$ is equal to a functional
$S[\vect{\phi}]$, which is the integral over the tiled domain $\Sigma$
of a function $\sigma(\nabla \vect{\phi})$ of the gradients of
$\vect{\phi}$~\cite{henley91}. The maximization of $S[\vect{\phi}]$
gives a (unique) function $\vect{\phi}^{(0)}$ which represents the
dominant tilings in the statistical ensemble under consideration.
The random tiling model states that $\sigma(\nabla \vect{\phi})$ has a
unique maximum, corresponding to tile fractions maximizing (in the
present case) the $2D$-fold rotational symmetry. The orientation of
the real space $\EC^{\parallel}$ is chosen so that the gradients are
zero at this maximum, and the model states that the entropy density
has a quadratic behavior near this maximum: $\sigma(\nabla
\vect{\phi}) = \sigma_{\mbox{\scriptsize max}} - {1 \over 2} \; \nabla
\vect{\phi} \cdot {\bf K} \cdot \nabla \vect{\phi} + o(|\nabla
\vect{\phi}|^2)$.  By analogy with an usual elastic theory, the tensor
{\bf K} is called the tensor of {\em phason elastic constants}. In the
basis of the perpendicular space $\EC^{\perp}$ where {\bf K} is
diagonal, the previous quadratic form becomes $\sigma (\nabla
\vect{\phi})= \sigma_{\mbox{\scriptsize max}} - {1 \over 2} \;
\sum_{i=0}^{D-3}\kappa_i (\nabla \phi_i)^2 + o(|\nabla
\vect{\phi}|^2)$.  We denote by $\vect{\theta}_i$ the normalized (in
$\EC^{\perp}$) eigenvector of {\bf K} corresponding to the eigenvalue
$\kappa_i$. We set
\begin{equation}
O_i[\vect{\phi}] = \int_{\Sigma} \vect{\phi}(r)\cdot\vect{\theta}_i \; d^2 r.
\label{Oi}
\end{equation}
$O_i[\vect{\phi}]$ is the mean
height of $\vect{\phi}$ in the direction $\vect{\theta}_i$.  As
written above, the total entropy of a function $\vect{\phi}$ is
$S[\vect{\phi}]=\int_{\Sigma} \sigma(\nabla \vect{\phi}(r)) \; d^2 r$.
To calculate the entropy $S(O_i)$ of the tilings of $\Sigma$ whose
mean height is $O_i$, we introduce the Lagrange multiplier $\lambda_i$
and we now maximize $G_i[\vect{\phi}] = S[\vect{\phi}] + \lambda_i
O_i[\vect{\phi}]$.  Without loss of generality we focus on $G_0$. Then
differentiating $G_0$ with respect to $\vect{\phi}$, we get, at first
order in $\vect{\phi}$ and its gradients, the equations $\frac{\delta
G_0}{\delta \phi_0} = \kappa_0 \Delta \phi_0 + \lambda_0 = 0$ and
$\frac{\delta G_0}{\delta \phi_i} = \kappa_0 \Delta \phi_i = 0$ for
$i>0$,  together with the condition that $\vect{\phi}$ matches the
boundary conditions induced by the tiling ones.  Let
$\vect{\phi}^{(0)}$ be the function maximizing $S[\vect{\phi}]$, which also
satisfies $O_0[\vect{\phi}^{(0)}]=0$, and $\vect{\phi}^{(1)} \equiv
\vect{\phi}-\vect{\phi}^{(0)}$.  By construction, $\vect{\phi}^{(1)}
\equiv 0$ at the boundary of $\Sigma$. The only non-zero coordinate of
$\vect{\phi}^{(1)}$ is $\phi^{(1)}_0$. It satisfies $\Delta
\phi^{(1)}_0 = - \lambda_0/\kappa_0$. At the large $D$ limit, $\Sigma$
tends towards a disk of radius $R =
pD/\pi$~\cite{largecodimI,largecodimII}. Taking into account the
condition $\phi^{(1)}_0 = 0$ on this circle, we get
$\phi^{(1)}_0(r) = - \frac{\lambda_0}{4 \kappa_0} (r^2-R^2)$.  Thus
$O_0[\vect{\phi}^{(1)}] = \displaystyle{-\frac{\pi}{8}
\frac{\lambda_0}{\kappa_0} R^4}$ and $\lambda_0 =
\displaystyle{-\frac{8 \kappa_0 O_0} {\pi R^4}}$.
One finally gets the relation (generalized to any $i$):
\begin{equation}
  S(O_i) = S[\vect{\phi}^{(0)}+\vect{\phi}^{(1)}] = S(0) - \frac{1}{2} 8 \pi^3 \kappa_i \left( \frac{O_i}{p^2D^2}
\right)^2
\end{equation}
for a diagonal tiling of side $p$ with (large) $2D$-fold symmetry.
Therefore at $T=\infty$, the equilibrium fluctuations of $O_i$ satisfy
$\langle O_i^2 \rangle = \frac{p^4 D^4}{8 \pi^3 \kappa_i}$
because of the Gaussian character of the number $\exp(S(O_i))$ of
tilings with mean height $O_i$. The measure of $\langle O_i^2 \rangle$
gives access to $ \kappa_i$.

Before giving our numerical results, we discuss the validity of this
calculation. It relies on the quadratic approximation above, which is
itself valid if $\nabla \vect{\phi}^{(0)} \cdot {\bf K} \cdot \nabla
\vect{\phi}^{(0)} \ll \sigma_{max} = O(1)$. Now $ | \nabla
\vect{\phi}^{(0)} |^2 =O(1/D)$ in finite $D$ tilings and phason
elastic constants are bounded \cite{largecodimI,largecodimII}. Thus
the quadratic approximation is valid at large $D$ and we get
asymptotically exact estimates of the $\kappa_i$. But at finite $D$,
this approach cannot be expected to provide accurate results. For
example, if $D=3$ one calculates that $| \nabla \vect{\phi}^{(0)} |^2$
can be as large as 4 in large macroscopic regions near the boundary,
and the previous argument collapses. We exemplify this procedure on
$\vect{\theta}_z = \frac{\sqrt{D-2}}{D} \sum_{\alpha=0}^{D-1}
(-1)^{\alpha} \vect{e}_{\alpha}$~\cite{henley91}. One checks that
$\| \vect{\theta}_z\|_{\EC^{\perp}}=1$. It is an eigenvector
of {\bf K} if and only if $D$ is even because it belongs to a
one-dimensional irreducible representation (irrep) in $H$ of the symmetry
group of the $2D$-gon, $C_{2Dv}$. This irrep
exists only if $D$ is even. $\vect{\theta}_z$ is a discrete component
of $\EC^{\perp}$ since for any $\alpha$, $\langle \vect{\theta}_z |
\vect{e}_{\alpha}^{\perp} \rangle_{\EC^{\perp}} = \pm 1/\sqrt{D-2}$
and any tiling vertex is represented in $\EC^{\perp}$ by a point of
coordinate multiple of $1/\sqrt{D-2}$ along $\vect{\theta}_z$.

Tab.~\ref{large:p:limits:kappa} gives the so-obtained values of
$\kappa_z$ in the large $p$ limit for $D=5,7,9,11$. The limits are
estimated by a fit of $\kappa_z$ at order 2 in $1/p$.
\begin{table}[ht]
\caption{Large $p$ estimates of the phason elastic constant $\kappa_z$.
\label{large:p:limits:kappa}}
$$
\begin{array}{|l|c|c|c|c|}
\hline
D & 5 & 7 & 9 & 11  \\
\hline
\kappa_z & \; 0.76 \; & \; 0.59 \; & \; 0.45 \; & \; 0.38 \; \\
\hline
\end{array}
$$
\end{table}
They clearly indicate that $\kappa_z \propto 1/D$ for the
values considered here. Thus our conclusion is that $\kappa_z$
vanishes at large $D$, proportionally to $1/D$, as predicted in
Ref.~\cite{largecodimI}. However the elastic constants studied there
were different from $\kappa_z$ since they were associated with
fluctuations of the de Bruijn lines separations, and existed whatever
the parity of $D$. This result is in agreement with the universality
of the large $D$ entropy per tile: since $ | \nabla \vect{\phi}^{(0)}
|^2 =O(1/D)$, the corrections to $\sigma_{\infty}$ vanish at large $D$
like $1/D^2$.

We have mentioned that this method cannot be expected to give reliable
results at finite $D$ and it is indeed what we observed for
$D=3$. However, for $D=5$, we already find, with the conventions of
normalization of~\cite{henley91}, $\kappa_z \simeq 0.26$. This value
is close to the expected one $\kappa_z = 0.29$~\cite{henley91}. In
principle it is possible to apply the same procedure to other
phason elastic constants $\kappa_i$. But determining the corresponding
$\vect{\theta}_i$ is a tedious task, requiring a complete
description of $\EC^{\perp}$ in terms of irreps
of $C_{2Dv}$. This goes beyond the scope of the present paper. 

\section{Conclusion} The computational power provided by the
Transition matrix Monte Carlo technique allowed us to investigate
random tilings with $2D$-fold rotational symmetry in their large size,
large $D$ limit. The numerical entropies obtained strongly support the
analytic predictions of previous
publications~\cite{largecodimI,largecodimII} and go into the direction
of a large $D$ ``universal'' entropy independent of size, shape, tile
fractions and boundary conditions. This work shows that the knowledge
of a few adjustable parameters that can be estimated by finite $D$ and
finite $p$ fits is sufficient to get a good estimate of finite $D$
entropies of physical interest (see Eq.~(\ref{generic})). For example,
the fact that the numerical value of $|e|$ is of the order of 1 shows
that the expansion of the finite $D$, infinite $p$, fixed-boundary
entropies at the first order in $1/D$ is accurate within few percents
as soon as $D \geq 4$. By contrast, the direct transposition of the
present results to the more physically relevant, finite $D$ free-boundary
tilings, is less immediate since it requires the numerical knowledge of
phason elastic constants.

We have demonstrated that the estimation of these phason
elastic constants for free-boundary, finite $D$ tilings is feasible
with a good accuracy. However, we did not tackle this task in the general case
but we have demonstrated its feasibility in a general framework. In the
case studied here, the phason elastic constant $\kappa_z$ falls off
like $1/D$, in agreement with earlier
predictions~\cite{largecodimI}. 

This work also illustrates that even
though polygonal boun\-daries are not physical~\cite{arctic3D},
fixed-boundary tilings are adapted to get valuable information on
their free- or periodic-boundary counterpart.  They present the great
advantage of being conveniently coded and manipulated in a computer
memory, in particular in relation with the {\em partition techniques}
used in several publications
(see~\cite{arctic3D,largecodimII,Desoutter05} and references therein).

\acknowledgments I express my gratitude to Mike Widom for helpful
discussions, comments and advice. I salute the participation of Suriya
Kothandaraman to the present work as part as his
Master's project.

\end{document}